\documentclass{article}
\usepackage{graphicx} % 
\usepackage{multicol}
\usepackage{xcolor}
\usepackage{verbatim}
\usepackage{amsmath,amssymb,amsthm,graphicx}
\usepackage{arxiv}
\usepackage{hyperref}
\newtheoremstyle{thm}% name
{9pt}%      Space above, empty = `usual value'
{9pt}%      Space below
{\itshape}% Body font
{}%         Indent amount (empty = no indent, \parindent = para indent)
{\bfseries}% Thm head font
{.}%        Punctuation after thm head
{ }% Space after thm head: \newline = linebreak
{}%         Thm head spec
\theoremstyle{thm}
\newtheorem{theorem}{Theorem}[section]

\newcommand{\R}{\mathbb{R}} % reelle
\newcommand{\N}{\mathbb{N}} % natuerliche
\usepackage{algorithm}
\usepackage[noend]{algpseudocode}
\makeatletter
\usepackage{arydshln}

\title{%Testing independence for circular data: kernel-based approach - Alternative: Negative definite kernel based tests of independence for spherical data
A new flexible class of kernel-based tests of independence
}

\author{ \href{https://orcid.org/0000-0001-5071-8350}{\includegraphics[scale=0.06]{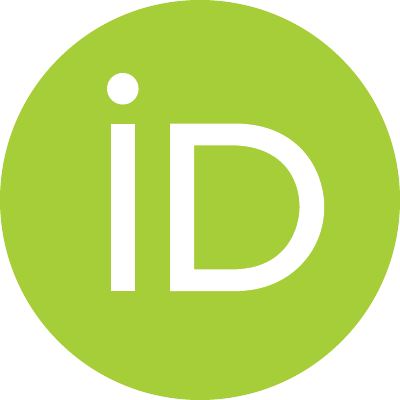}\hspace{1mm} Marija Cupari\' c} \\
	University of Belgrade\\
	Faculty of Mathematics\\
	Belgrade, 11000, Serbia \\
	\texttt{marija.cuparic@matf.bg.ac.rs} \\
	\And
\href{https://orcid.org/0000-0003-4329-8794}{\includegraphics[scale=0.06]{orcid.pdf}\hspace{1mm}Bruno Ebner} \\
	Institute of Stochastics \\ Karlsruhe Institute of Technology (KIT)\\
	Englerstr. 2, 76131 Karlsruhe, Germany \\
	\texttt{bruno.ebner@kit.edu} \\
 \And
    	\href{https://orcid.org/0000-0001-8243-9794}{\includegraphics[scale=0.06]{orcid.pdf}\hspace{1mm}Bojana Milo\v sevi\' c} \\
	University of Belgrade\\
        Faculty of Mathematics\\
	Belgrade, 11000, Serbia \\
	\texttt{bojana.milosevic@matf.bg.ac.rs} \\
}

\date{\today}

\begin{document}

\maketitle

\begin{abstract}
    Spherical and hyperspherical data are commonly encountered in diverse applied research domains, underscoring the vital task of assessing independence within such data structures. In this context, we investigate the properties of test statistics relying on distance correlation measures originally introduced for the energy distance, and  generalize the concept to strongly negative definite kernel-based distances. An important benefit of employing this method lies in its versatility across diverse forms of directional data, enabling the examination of independence among vectors of varying types. The applicability of tests is demonstrated on several real datasets.
\end{abstract}

\section{Introduction}
In this paper we apply the kernel distance correlation concept to circular, spherical and hyperspherical data. Measures of directional dependence have been widely explored in the literature; see \cite{MJ:2000} chapter 11 and the references therein, including distribution-free rank-based methods. As an example, classical correlation measures for circular-circular, hence toroidal, data include the correlation coefficients of Jammalamadaka-Sarma, see \cite{jammalamadaka1988correlation}, or Fisher-Lee, see \cite{fisher1983correlation}, which are designed to mimic the ordinary correlation coefficient, see also \cite{Chakraborty2023}.  In contrast to these, the hitherto literature lacks the distance correlation concept for directional data, a gap that we try to explore by defining the distance correlation on the sphere. Distance correlation was first proposed in \cite{szekely2005hierarchical} and developed in \cite{SRB:2007,szekely2009brownian,szekely2014partial}, where one finds the definition of the so called population distance covariance, see also \cite{lyons2013distance} for a related generalization in metric spaces. We refer to \cite{Edelmann_2021} for a comprehensive overview and for more details and several examples of applications to the references therein. The goal is to propose new independence tests for directional-directional data. 

To be specific, denote with $\|\cdot\|$ the usual Euclidean distance and let $\mathcal{S}^{d_1}=\{x\in\R^{d_1}:\,\|x\|=1\}$ and $\mathcal{S}^{d_2}$ be two spheres of dimensions $d_1,d_2\in\N$. Let $(X,Y)$ be a pair of random elements in $\mathcal{S}^{d_1}\times\mathcal{S}^{d_2}$. Denote with $(X',Y')$, $(X'',Y'')$ two independent copies of $(X,Y)$. We define the kernel-induced population distance covariance of $X$ and $Y$ 
\begin{align}\label{eq:dCov}
\mbox{dCov}^2_K(X,Y)&=E\left[K(X,X')K(Y,Y')\right]+E\left[K(X,X')\right]E\left[K(Y,Y')\right]-2E\left[K(X,X')K(Y,Y'')\right]
\end{align}
and the kernel induced population distance correlation to be
\begin{equation}\label{eq:dCor}
    \mbox{dCor}_K(X,Y)=\frac{\mbox{dCov}_K(X,Y)}{\sqrt{\mbox{dCov}_K(X,X)\mbox{dCov}_K(Y,Y)}}.
\end{equation}
%Note that we assume that $d_j$, $j=1,2$, are of strong negative type to ensure consistency of estimators, for details see \cite{lyons2013distance}.
Note that we assume the kernels $K$ to be strongly negative definite, and that we get the so called population Hilbert-Schmidt covariance for the particular choice of a Gaussian kernel, see \cite{sejdinovic2013equivalence}. Notice that since $\mathcal{S}^{d_j}\subset\R^{d_j}$ and putting $K(x,y)=\|x-y\|^a$, $a\in(0,2)$, the so-called generalized energy kernel, we have the distance correlation by Sz\'{e}kely and Rizzo, for details on this topic, see \cite{szekely2023energy}. The idea of generalization in \eqref{eq:dCov} compared to the distance covariance is to replace the underlying energy distance as a distance of probability measures of Sz\'{e}kely and Rizzo by the kernel-induced distances of the probability measures as defined in Appendix \ref{app:Kernel} inspired by \cite{R:2013}, Section 22 and \cite{K:2006}.

The novelty of this paper is to choose different types of kernels $K$ and to consider the induced distance correlation measures. Since the dimensions $d_1$ and $d_2$ do not have to coincide, we are able to consider correlations between (say) spherical and circular, spherical and hyperspherical, or even spherical and linear random elements. The problem of interest is in each of these settings to test
\begin{equation*}
H_0:\; X\,\mbox{and}\,Y\, \mbox{are independent}
\end{equation*}
against general alternatives. Distance correlation as defined in \eqref{eq:dCor} is especially suitable for designing a test for this testing problem, since $X$ and $Y$ are independent if and only if dCor$_K(X,Y)$=0.

 Clearly, we need a consistent estimator of \eqref{eq:dCov} to apply the concept to the data. Based on an independent and identically distributed (i.i.d.) sample $\{(X_i,Y_i)\}$, $j=1,\ldots,n$, distributed as $(X,Y)$, we consider a test statistic
\begin{align*}
    \Upsilon^2_n=\frac{V_{n,X,Y}}{\sqrt{V_{n,X,X}}\sqrt{V_{n,Y,Y}}},
\end{align*}
where 
\begin{align}
    V_{n,X,Y}&=\frac{1}{n^2}\!\!\sum_{i,j=1}^nK(X_i,X_j)K(Y_i,Y_j)+\frac{1}{n^4} \!\!\sum_{i,j=1}^n\!\!K(X_i,X_j)\sum_{i,j=1}^n\!\!K(Y_i,Y_j)-\frac{2}{n^3}\!\!\sum_{i,j,k=1}^n\!\!K(X_i,X_j)K(Y_i,Y_k).
\end{align}
Since the test statistics should be close to zero under independence, we reject the hypothesis for large values of $\Upsilon^2_n$.

Beside the generalized energy kernel, we also consider the kernels 
\begin{equation*}
K_k(x,y)=\frac{||x-y||}{1+||x-y||}\quad\mbox{and}\quad K_l(x,y)=\log(1+||x-y||^2),
\end{equation*}
which are all of strong negative type, see p. 600 in \cite{R:2013}. We denote the corresponding test statistics with $D_{a,n}$ $D_{k,n}$ and $D_{l,n}$. 

The rest of the paper is structured as follows. In Section \ref{sec:Asymp} we provide the asymptotic null distribution of the test statistics using $V$-statistic theory, in Section \ref{sec:power} we provide a comparative Monte Carlo simulation study for the case of toroidal data, for spherical-spherical data and for the circular-linear case. Three real data examples are provided in Section \ref{sec:real} and we finish the paper by providing some concluding remarks and an outlook for further research as well as an Appendix on kernel induced probability metrics.

\section{Asymptotic properties}\label{sec:Asymp}
{
Let $(X_1,Y_1),..., (X_n,Y_n)$ be an i.i.d. random sample from  $\mathcal{S}^{d_1}\times \mathcal{S}^{d_2}$, and the kernel function $K$ satisfies previously mentioned conditions. In order to obtain the limiting null distribution of the test statistic $\Upsilon^2_n$
 we first    consider the limiting properties of $V_{n,X,Y}.$
 It is clear that $V_{n,X,Y}$ is a $V$-statistic of order 4, with kernel  
 \begin{align*}
    \Phi((x_1,y_1),(x_2,y_2),(x_3,y_3),(x_4,y_4))&=K(x_1,x_2)K(y_1,y_2)+K(x_1,x_2)K(y_3,y_4)\\&-2K(x_1,x_2)K(y_1,y_3).
\end{align*}
 The first projection of its symetrized version, is
\begin{align*}
    \varphi_1((x_1,y_1))&=E(\Phi((X_1,Y_1),(X_2,Y_2),(X_3,Y_3),(X_4,Y_4))|(X_1,Y_1)=(x_1,y_1))\\&=\frac{1}{2}\Big(E(K(X_2,X_3)K(Y_2,Y_3))+E(K(x_1,X_2)K(y_1,Y_2))+E(K(x_1,X_2)K(Y_3,Y_4))\\&+E(K(X_3,X_4)K(y_1,Y_2))-E(K(x_1,X_2)K(y_1,Y_3))-E(K(X_2,x_1)K(Y_2,Y_4))\\&-E(K(X_2,X_3)K(Y_2,y_1))-E(K(X_2,X_3)K(Y_2,Y_4))\Big).
\end{align*}

Under $H_0$, it can be seen that $ \varphi_1((x_1,y_1))=0$, i.e. the $V$-statistic is degenerate. Therefore, using the theory of $V$-statistics (see \cite{korolyuk}), the limiting null distribution is determined by the second projection of the kernel $\Phi$ which is equal to
\begin{align}\label{drugaProjekcija}
\begin{aligned}
    \varphi_2((x_1,y_1),(x_2,y_2))&=\frac{1}{6}\Big(EK(x_2,X_3)EK(y_2,Y_3)+EK(x_1,X_3)EK(y_1,Y_3)\\&+EK(x_1,X_3)EK(y_2,Y_3)+ EK(x_2,X_3)EK(y_1,Y_3)\\&+EK(X_3,X_4)EK(Y_3,Y_4)+K(x_1,y_1)K(x_2,y_2)\\&+K(x_1,x_2)EK(Y_3,Y_4)+K(y_1,y_2)EK(X_3,X_4)\\&-K(x_1,x_2)EK(y_1,Y_3)-K(x_1,x_2)EK(y_2,Y_3)\\&-K(y_1,y_2)EK(x_1,X_3)-K(y_1,y_2)EK(x_2,X_3)\\&-EK(X_3,x_1)EK(Y_3,Y_4)-EK(X_3,x_2)EK(Y_3,Y_4)\\&-EK(X_3,X_4)EK(Y_4,y_1)-EK(X_3,X_4)EK(Y_3,y_2)\Big).
    \end{aligned}
\end{align}
Hence, we have under $H_0$ the convergence in distribution
$nV_n\overset{d}{\to}6\sum_{i=1}^{\infty}\eta_iW_i^2$ for $n\rightarrow\infty$, where $\{W_i\}$ is a sequence of i.i.d. standard normal random variables and $\{\eta_i\}$ is the positive sequence of eigenvalues of an integral operator with kernel $\varphi_2((x_1,y_1),(x_2,y_2))$ i.e. the solutions of the integral equation
\begin{align}\label{operator}
 \eta q(x_1,y_1)=\int_{\mathcal{S}^{d_1}}\int_{\mathcal{S}^{d_2}}\varphi_2((x_1,y_1),(x_2,y_2))q(x_2,y_2)f_X(x_2)f_Y(y_2)dx_2dy_2.   
\end{align}

}
%In order to explore the asymptotic properties of $D_{a,n},D_{k,n}$ and $D_{l,n}$ we consider
%\scalebox{0.8}{
%\begin{tabular}{c|cc|cc}
%   &\multicolumn{2}{c|}{$n=20$}& \multicolumn{2}{c}{$n=50$} \\
%     Model  & $C_n$ &$D_n$& $C_n$ &$D_n$\\ \hline
%$X\sim VM(0,1)$, $Y\sim VM(\pi,0.1)$ & 4 & 8 & 5 & 5 \\
%$X\sim VM(0,1)$, $Y\sim VM(0,0.1)$ & 5 & 6 & 5 & 6\\
%$X\sim VC(0,exp(-0.1))$, $Y\sim VC(\pi,exp(-0.1))$ & 6 & 7 & 5 & 7 \\
%$X\sim VC(0,exp(-0.1))$, $Y\sim VC(0,exp(-0.1))$ & 6 & 8 & 4 & 6\\
%     \hline
%     $X\sim VM(0,1)$, $Y=X$, with probability  0.5, otherwise $Y\sim VM(\pi,0.1)$ & 46 & 72 & 84 & 98 \\
%      $X\sim VM(0,1)$, $Y=X$, with probability  0.2, otherwise $Y\sim VM(\pi,0.1)$ & 11 & 19 & 19 & 32\\
%     $X\sim VM(0,3)$, $Y=atan2(0.15\cos(x) + 0.25\sin(x), 0.35\sin(x))+VM(0,1)$ & 40 & 52 & 84 & 92\\
%     $X\sim VC(0,exp(-0.1))$, $Y=X$, with probability  0.5, otherwise $Y\sim VC(\pi,exp(-0.1))$ &  54 & 76 & 92 & 99\\
%\end{tabular}}
%a $V-$statistic $V_{n,X,Y}$ defined as
%\begin{align*}
 %   V_{n,X,Y}&=\frac{1}{n^2}\sum\limits_{i,j}K(X_i,X_j)K(Y_i,Y_j)+\frac{1}{n^4}\sum\limits_{i,j}K(X_i,X_j)\sum\limits_{i,j}K(Y_i,Y_j)-\frac{2}{n^3}\sum\limits_{i,j,k}K(X_i,X_j)K(Y_i,Y_k)\\&=\frac{1}{n^4}\sum\limits_{i,j,k,l}\Phi((X_i,Y_i),(X_j,Y_j),(X_k,Y_k),(X_l,Y_l)).
%\end{align*}

Next, % notice that each of $D_{a,n},D_{k,n}$ and $D_{l,n}$, labeled for brevity by $\Upsilon_n$, can  be represented as a quotient 
since 
\begin{align*}
    \Upsilon^2_n=\frac{V_{n,X,Y}}{\sqrt{V_{n,X,X}}\sqrt{V_{n,Y,Y}}},
\end{align*}
 and $V_{n,X,X}$ and  $V_{n,Y,Y}$, being one-sample V-statistics, converge to their expectations %given by 
%$\sigma^2_{K,X}=EK^2(X_1,X_2)+(EK(X_1,X_2))^2-%2EK(X_1,X_2)K(X_1,X_3)$ and %$\sigma^2_{K,Y}=EK^2(Y_1,Y_2)+(EK(Y_1,Y_2))^2-%2EK(Y_1,Y_2)K(Y_1,Y_3)$,
by applying Slutsky's theorem we get that, under $H_0$ we get the statement of the following theorem.

\begin{theorem}
  Let $(X_1,Y_1),..., (X_n,Y_n)$ be an i.i.d. random sample from  $\mathbb{X}\times \mathbb{Y}$, such that  $X_1$ and $Y_1$ are independent, and the kernel function $K$ satisfies previously mentioned conditions and $EK^2(X_1,Y_1)<\infty$. Then
   $$n\Upsilon_n^2\overset{d}{\to} \frac{6}{\sigma_{K,X}\sigma_{K,Y}}\sum_{i=1}^{\infty}\eta_iW_i^2,$$
   where $\{\eta_i\}$ is the sequence of eigenvalues of the integral operator with kernel \eqref{drugaProjekcija} and $\{W_i\}$ the sequence of i.i.d. random variables with standarad noraml distribution, and $\sigma^2_{K,X}=EK^2(X_1,X_2)+(EK(X_1,X_2))^2-2EK(X_1,X_2)K(X_1,X_3)$ and $\sigma^2_{K,Y}=EK^2(Y_1,Y_2)+(EK(Y_1,Y_2))^2-2EK(Y_1,Y_2)K(Y_1,Y_3)$.
\end{theorem}

As a consequence of the reasoning above we see that under alternatives, under the same assumptions for the function $K$ (including second-moment finiteness), $\Upsilon^2_n$ converges for $n\rightarrow\infty$ almost surely to a positive constant, implying that $n\Upsilon^2_n\rightarrow\infty$ almost surely for $n\rightarrow\infty$. This shows that the test is consistent against all alternatives. %\textcolor{blue}{Do we need a moment assumption under the alternatives here?}

Since $\eta_i$ are solutions of \eqref{operator}, and depends on marginal distributions $F_{X}$ and $F_Y$, the limiting distribution of $n\Upsilon_n^2$ is not distribution-free. One way to overcome this is to use an appropriate resampling procedure. 

Here we might employ the following permutation bootstrap ($\Upsilon_n$ labels the statistic of interest, see e.g. \cite{garcia2014test}). %In the case of computationally demanding simulations

\begin{algorithm}[h!]
%\captionsetup{labelformat=empty}
\caption{\bf Bootstrap  algorithm }
\label{bootstrap}
\begin{algorithmic}[1]
    \State  Based on $(X_1,Y_1),...,(X_n,Y_n)$ compute the test statistic $\Upsilon_n$;
 %  \State  Estimate critical value $q^*_{n,1-\alpha}$: 
   
    \For{$i$ from $1$ to $B$}
        \State   construct bootstrap sample $(X^*_1,Y^*_1),...,(X^*_n,Y^*_n)$, where $X^*_i=X_i$ and  $Y^*_i=Y_{\sigma(i)}$ for some random permutation $\sigma:\{1,...,n\}\mapsto\{1,...,n\}$;  
  \State determine the value of the test statistic
        \vspace*{-3mm}
        $$\Upsilon^*_{n,i}=\Upsilon_{n,i}((X^*_1,Y^*_1),...,(X^*_n,Y^*_n));$$
     %   \State (e) repeat steps (a)-(d)  B times;
    \EndFor  
    \State Based on $\Upsilon^*_{n,1},\!...,\Upsilon^*_{n,B}$ estimate critical region $W^*_{n,\alpha}$;
    \If{$\Upsilon_n\in W^*_{n,\alpha}$} 
        \State Reject $H_0$.
    \EndIf
\end{algorithmic}
\end{algorithm}

\section{Power study}\label{sec:power}
In this section we evaluate the power performance of the novel tests $D_{a,n},D_{k,n}$ and $D_{l,n}$
that corresponds to $\Upsilon_n$ for the strongly negative kernels
\begin{itemize}
    \item $K_a(x,y)=||x-y||^a,\;a\in(0,2);$
    \item $K_k(x,y)=\frac{||x-y||}{1+||x-y||};$
    \item $K_l(x,y)=\log(1+||x-y||^2).$
\end{itemize}

Given that our class of tests is rather flexible, we explore the small-sample properties for data on the torus (circular-circular data), for spherical-spherical data, and in the case of spherical-linear data. In all cases powers are estimated for the level of significance $\alpha=5\%$ and sample sizes $n=20$ and $n=50$, and using the bootstrap algorithm \ref{bootstrap} with $N=2000$ Monte Carlo replicates and $B=1000$ bootstrap cycles. In the case of circular-linear data, due to computational complexity of a competitor, we use the warp-speed modification (see \cite{giacomini2013warp}) with $N=2000$ Monte Carlo replicates.
The competitors and alternatives are stated for each case separately.

\subsection{Case I: toroidal data}
%as well as the following competitors:
In order to unify notation with those used in the literature, we use angular representation, i.e, denote by $(\theta_i^{(1)},\theta_i^{(2)})$  the pairs of angles that corresponds to $(X_i,Y_i)$. 

Due to the lack of competitors in the literature, we only consider

 \begin{itemize}
     \item $C_n$ the circular correlation coefficient which is defined as the empirical counterpart of 
\begin{align*}
    cCor(\theta^{(1)},\theta^{(2)})=\frac{E(\sin(\theta^{(1)}-\mu)\sin(\theta^{(2)}-\nu))}{\sqrt{Var(\sin(\theta^{(1)}-\mu))Var(\sin(\theta^{(2)}-\nu))}},    
\end{align*}
where $\mu$ and $\nu$ are the mean directions of $\theta^{(1)}$ and $\theta^{(2)}$, respectively. This test is studied in \cite{jammalamadaka1988correlation} and implemented in the R package circular (see \cite{Therneau2020-xf});
\item  a test based on trigonometric moments proposed in \cite{garcia2023nonparametric} with the test statistic
\begin{align*}
   T_{n,\lambda}&=\frac{1}{n}\sum_{j,k}\mathcal{J}_{c}^{(v)}(\theta_{jk}^{(1)})\mathcal{J}_{c}^{(v)}(\theta_{jk}^{(2)})+\frac{1}{n^3}\Big(\sum_{j,k}\mathcal{J}_{c}^{(v)}(\theta_{jk}^{(1)})\Big)\Big(\sum_{j,k}\mathcal{J}_{c}^{(v)}(\theta_{jk}^{2})\Big)\\&-\frac{2}{n^2}\sum_{j,k,l}\mathcal{J}_{c}^{(v)}(\theta_{jk}^{(1)})\mathcal{J}_{c}^{(v)}(\theta_{jl}^{(2)}), 
\end{align*}
where
\begin{align*}
    \mathcal{J}_{c}^{(v)}(\theta)=\cos(\lambda \sin(\theta))e^{\lambda(\cos(\theta)-1)},\;\lambda>0\\
    \theta_{jk}^{(m)}=\theta_j^{(m)}-\theta_k^{(m)},\; j,k=1..n,\;m\in\{1,2\}.\\
\end{align*}
Having in mind the recommendations of authors in the original paper, we consider the test for $\lambda=1.$
 \end{itemize}

The families of bivariate circular distributions in order to check the size of the tests are denoted by $F\times G$ where $F$ and $G$ are independent univariate circular distributions that belong to one of the following families:
\begin{itemize}
    \item  Von Mises distribution $VM(\mu,\kappa)$ with a density
\begin{align}\label{univariateVonMises}
        f(\theta)=\frac{1}{2\pi\mathcal{I}_0(\kappa)}e^{\kappa\cos(\theta-\mu)},\;\theta\in[0,2\pi),
    \end{align}
    where $\mathcal{I}_r(\cdot)$ denotes the modified Bessel function of the first kind of order $r$, and $0\leq\mu <2\pi$ and $\kappa\geq0$ are the location and concentration parameters, respectively.
    \item  Wrapped Cauchy distribution $WC(\mu,\rho)$ with a density
    \begin{align}\label{wrapped}
        f(\theta)=\frac{1}{2\pi}\frac{1-\rho^2}{1+\rho^2-2\rho\cos(\theta-\mu)},\;\theta\in[0,2\pi),
    \end{align}
    where $\mu\in[0,2\pi)$ is a location parameter, and $\rho\in[0,1)$ controls the concentration of the model.
\end{itemize}
%as well as some of the following distributions (that were also used for studying power performance)
For assessing both test sizes and powers we consider 
\begin{itemize}
    \item the parabolic distribution $PB(p)$ defined by
    $\theta^{(1)}\sim Unif([0,2\pi))$ and $\theta^{(2)}=2(p(\theta^{(1)})^2+(1-p)U^2)$, where $U\sim Unif([0,2\pi])$  and is independent of $\theta^{(1)}$, $p\in[0,1]$ (see \cite{garcia2023nonparametric}).
    \item The centered Bivariate Von Mises distribution proposed by \cite{shieh2005inferences} $BvM(\kappa_1,\kappa_2,\mu_g,\kappa_g)$ and with density 
    \begin{align}\label{bivariate}
        f(\theta^{(1)},\theta^{(2)})=f_1(\theta^{(1)})f_2(\theta^{(2)})f_g(2\pi(F_1(\theta^{(1)})-F_2(\theta^{(2)}))),
    \end{align}
    where $f_j$ and $F_j$ are respectively the marginal densities and distribution functions corresponding to $VM(0,\kappa_j),\;j=1,2$. Here, $f_g$ is a density function that corresponds to $VM(\mu_g,\kappa_g),\;\mu_g\in[0,2\pi],\;\kappa_g\geq 0$.
\item The centered Bivariate Wrapped Cauchy distribution $BWC(\rho_1,\rho_2,\rho)$ (see \cite{pewsey2016parametric}) with the density of the form \eqref{bivariate}, where
 $f_j$ and $F_j$ are respectively the marginal densities and distribution functions corresponding to $WC(0,\rho_j),\;j=1,2,\;\rho_j\in[0,1)]$  and $f_g$ is a density function that corresponds to $WC(0,|\rho|),$ and $\rho\in(-1,1)$.
    
    \item The centered Bivariate Cosine von Mises distribution with positive interaction $BCvM(\kappa_1,\kappa_2,\kappa_3)$ with density
    \begin{align*}
        f(\theta^{(1)},\theta^{(2)})=C(\kappa_1,\kappa_2,\kappa_3)\exp(\kappa_1\cos(\theta^{(1)})+\kappa_2\cos(\theta^{(2)})+\kappa_3\cos(\theta^{(1)}-\theta^{(2)})),
    \end{align*}
    where $\kappa_1,\kappa_2\geq 0$ are concentration parameters and $\kappa_3\in\mathbf{R}$ is a parameter controlling dependence and $C(\kappa_1,\kappa_2,\kappa_3)$  normalizing constant.
    \item A mixture type distribution $Mix(F,G,p)$ where $(\theta^{(1)},\theta^{(2)})$ are defined by
    $\theta^{(1)}\sim F$ and $\theta^{(2)}=\theta^{(1)}$ with probability $p$ or $\theta^{(3)}$ which is independent from $\theta^{(1)}$ and has distribution $G$.
\end{itemize}

The results for $n=20$ and $n=50$ are presented in Tables \ref{tab: circular20} and \ref{tab: circular50}. It is important to notice that all tests are well calibrated with few exceptions with slightly liberal behaviour. Therefore, their comparison is fully justified. We can see that the traditional circular-correlation coefficient-based test is almost always less powerful than the other tests. In the case of energy kernel, the parameter $a$ seams not to have significant impact, hence we suggest using this test with $a=1$, while if there is a suspicion on a mixture distribution, smaller values of $a$ are more convenient. 
The results for other kernels are similar, but the ratio-based kernel statistics $D_r$ perform slightly better for the alternatives considered.

\begin{table}[t]
    \centering

\scalebox{0.7}{
\begin{tabular}{c|ccccccccccc}
 %  &\multicolumn{11}{c}{$n=20$} %& \multicolumn{8}{c}{$n=50$}    \\
     Model  & $C$ & $T_1$ & $D_{0.25}$ & $D_{0.5}$ & $D_{0.75}$ & $D_{1}$ & $D_{1.25}$ 
     & $D_{1.5}$ & $D_{1.75}$& $D_k$ & $D_l$ %& C & $D_{0.25}$ & $D_{0.5}$ & $D_{0.75}$ & $D_{1}$ & $D_{1.25}$      & $D_{1.5}$ & $D_{1.75}$
     \\ \hline
     
$VM(0,1)\times VM(\pi,0.1)$ & 5 & 6 &  6 & 5 & 5 & 5 & 5 & 5 & 5 & 6 & 5   \\
$ VM(0,1)\times VM(0,0.1)$ & 6 & 5 &  5 & 5 & 5 & 5 & 5 & 5 & 5 & 5 & 5  \\
$WC(0,exp(-0.1))\times WC(\pi,exp(-0.1))$ & 6 & 5 & 5 & 5 & 5 & 5 &  5 & 5 & 5 & 5 & 5 \\
$WC(0,exp(-0.1))\times WC(0,exp(-0.1))$ & 5 &  6 & 6 & 6 & 6 & 6 & 6 & 6 & 6 & 6 & 6 \\
$PB(0)$ & 7 & 6 & 6 & 6 & 6 & 6 & 5 & 6 & 6 & 6 & 6\\
$BvM(0)$ & 5 & 5 &  5 & 5 & 5 & 5 & 5 & 5 & 5 & 5 & 5 \\
$BWC(0)$ & 5 & 5 & 5 & 5 & 5 & 5 & 5 & 5 & 5 & 5 & 5 \\
$BCvM(0)$ & 4 & 4 & 5 & 5 & 5 & 5 & 5 & 4 & 4 & 5 & 5 \\

     \hline
     $Mix(VM(0,1),VM(\pi,0.1),0.5)$ & 44 & 72 & 73 & 72 & 70 & 68 & 66 & 66 & 65 & 73 & 67   \\
      $Mix(VM(0,1),VM(\pi,0.1),0.2)$& 9 & 14 & 14 & 14 & 14 & 13 & 13 & 14 & 14 & 15 & 14   \\
    % $X\sim VM(0,3)$, $Y=atan2(0.15\cos(x) + 0.25\sin(x), 0.35\sin(x))+VM(0,1)$ & 0.4160 & 0.4150 & 0.4055 & 0.4475 & 0.4745 & 0.4910 & 0.4890 & 0.4885 & 0.4885 & 0.4265 & 0.4850   \\
    $Mix(VC(0,exp(-0.1)),VC(\pi,exp(-0.1)),0.5)$  & 54 & 73 & 73 & 73 & 72 & 71 & 70 & 69 & 69 & 74 & 70   \\
     $PB(0.2)$ & 8 & 12 & 12 & 12 & 11 & 10 & 9 & 8 & 8 & 12 & 9  \\
     $PB(0.4)$ & 24 &37 &  36 & 37 & 37 & 37 & 36 & 36 & 35 & 37 & 36  \\
     $PB(0.6)$ & 52 & 87 & 86 & 89 & 91 & 91 & 92 & 92 & 92 & 87 & 91 \\
     $PB(0.8)$ & 46 & 100 & 100 & 100 & 100 & 100 & 100 & 100 & 100 & 100 & 100 \\
     $PB(1)$ & 27 & 100 & 100 & 100 & 100 & 100 & 100 & 99 & 97 & 100 & 100  \\
     $BvM(0.5)$ & 14 & 19 & 17 & 19 & 19 & 19 & 19 & 19 & 18 & 18 & 19 \\
     $BvM(1)$ & 36 & 58 &  53 & 56 & 57 & 56 & 54 & 52 & 50 & 55 & 55 \\
     $BvM(1.5)$ & 60 & 89 & 86 & 88 & 88 & 86 & 85 & 83 & 80 & 88 & 86 \\
     $BvM(2)$ & 76 & 98 & 97 & 98 & 98 & 98 & 97 & 96  & 95 & 98 & 98 \\
     $BvM(3)$ & 90 &  100 &  100 & 100 & 100 & 100 & 100 & 100 & 100 & 100 & 100 \\
$BWC(-0.2)$ & 11 & 13 & 12 & 12 & 13 &  13 & 13 & 14 & 14 & 12 & 13  \\
     $BWC(-0.4)$ &  28 &  43 & 42 & 45 & 47 & 48 & 48 & 49 & 49 & 42 & 48\\
     $BWC(-0.6)$ & 57 & 87 & 85 & 88 & 89 &  89 & 90 & 90 & 90 & 86 & 90 \\
    $BWC(-0.8)$ & 83 & 100 & 99 & 100 & 100 & 100 & 100 & 100 & 100 & 100 & 100 \\
     
   $ BCvM(0.5)$ & 15 & 12 & 11 & 12 & 14 & 15 & 15 & 16 & 16 & 11 & 15  \\
     $BCvM(1)$ & 39 & 32 & 29 & 34 & 38 & 41 & 42 & 44 & 45 & 31 & 42 \\
     $BCvM(1.5)$ & 67 & 58 &  52 & 61 & 66 & 70 & 72 & 73 & 74 & 55 & 71 \\
     $BCvM(2)$ & 85 & 77 & 70 & 78 & 83 & 85 & 87 & 89 & 90 & 73 & 87  \\
    $BCvM(3)$ & 97 & 94 & 88 & 94 & 96 & 97 & 98 & 98 & 98 & 91 & 98 \\
\end{tabular}}
    \caption{Empirical sizes and powers of tests in Case I ($n=20$, $\alpha=0.05$)}
    \label{tab: circular20}
\end{table}

\begin{table}[t]
    \centering

\scalebox{0.7}{
\begin{tabular}{c|ccccccccccc}
 %  &\multicolumn{11}{c}{$n=50$} %& \multicolumn{8}{c}{$n=50$} 
   \\
     Model  & $C_d$ & $T_1$ & $D_{0.25}$ & $D_{0.5}$ & $D_{0.75}$ & $D_{1}$ & $D_{1.25}$ 
     & $D_{1.5}$ & $D_{1.75}$& $D_k$ & $D_l$ %& C & $D_{0.25}$ & $D_{0.5}$ & $D_{0.75}$ & $D_{1}$ & $D_{1.25}$      & $D_{1.5}$ & $D_{1.75}$
     \\ \hline
     
$VM(0,1)\times VM(\pi,0.1)$ & 5 & 6 &  5 & 5 & 6 & 6 & 6 & 6 & 6 & 6 & 6  \\
$VM(0,1)\times VM(0,0.1)$ & 6 & 5 & 6 & 6 & 6 & 6 & 5 & 5 & 5 & 6 & 5  \\
$WC(0,exp(-0.1))\times WC(\pi,exp(-0.1))$ & 5 & 5 & 5 & 5 & 5 & 6 & 5 & 5 & 5 & 5 & 5 \\
$WC(0,exp(-0.1))\times WC(0,exp(-0.1))$ & 5 & 6 & 5 & 6 & 6 & 6 & 5 & 5 & 5 & 5 & 6  \\
$PB(0)$ & 5 & 5 & 6 & 5 & 6 & 5 & 5 & 5 & 5 & 5 & 5 \\
 $BvM(0)$ & 4 & 5 & 5 & 5 & 5 & 4 & 4 & 4 & 4 & 5 & 5 \\
 $BWC$ &  5 & 6 & 7 & 6 & 6 & 6 & 6 & 6 & 5 & 6 & 6 \\
 $BCvM(0)$ & 6 & 5 & 5 & 5 & 5 & 5 & 5 & 5 & 5 & 5 & 5  \\
     \hline
      $Mix(VM(0,1),VM(\pi,0.1),0.5)$ & 83 & 99 & 99 & 99 & 99 & 98 & 98 & 97 & 97 & 99 & 98    \\
      $Mix(VM(0,1),VM(\pi,0.1),0.2)$& 17 & 31 & 33 & 30 & 28 & 28 & 27 & 26 & 26 & 32 & 28   \\
     %$X\sim VM(0,3)$, $Y=atan2(0.15\cos(x) + 0.25\sin(x), 0.35\sin(x))+VM(0,1)$ & 0.8230 &  & 0.8285 & 0.8740 & 0.8915 & 0.8975 & 0.9020 & 0.8970 & 0.8910 & 0.8505 & 0.8970 \\
     $Mix(VC(0,exp(-0.1)),VC(\pi,exp(-0.1)),0.5)$ & 92 & 99 & 99 & 99 & 99 & 98 & 98 & 98 & 98 & 99 & 98   \\
     $PB(0.2)$ & 12 & 24 & 28 & 24 & 21 & 18 & 16 & 14 & 13 & 28 & 17 \\ 
     $PB(0.4)$ & 51 & 86 & 86 & 86 & 85 & 84 & 81 & 79 & 77 & 87 & 83 \\
     $PB(0.6)$ & 73 & 100 & 100 & 100 & 100 &  100 & 100 & 100 & 100 & 100 & 100 \\
     $PB(0.8)$ & 64 & 100 & 100 & 100 & 100 & 100 & 100 & 100 & 100 & 100 & 100  \\
     $PB(1) $& 30 & 100 & 100 & 100 & 100 & 100 & 100 & 100 & 100 & 100 &  100 \\
     $BvM(0.5)$ & 29 & 44 & 41 & 44 & 43 & 42 & 41 & 40 & 37 & 42 & 42  \\
     $BvM(1)$ & 73 & 96 & 94 & 95 & 95 & 94 & 93 & 92 & 90 & 95 & 94  \\
     $BvM(1.5)$ &  94 & 100 & 100 & 100 & 100 & 100 & 100 & 100 & 100 & 100 & 100\\
     $BvM(2)$ & 99 & 100 & 100 & 100 & 100 & 100 & 100 & 100 & 100 &  100 & 100 \\
     $BvM(3)$ & 100 &  100 & 100 & 100 & 100 & 100 & 100 & 100 & 100 & 100 & 100 \\
$BWC(-0.2)$ & 20 & 28 & 27 & 29 & 30 &  31 & 32 & 32 & 32 & 28 & 32 \\
      $BWC(-0.4)$ & 55 & 87 & 86 & 88 & 90 & 90 & 90 & 91 & 91 & 86 & 90 \\
      $BWC(-0.6)$ &  81 & 100 & 100 & 100 & 100 & 100 & 100  &100 & 100 & 100 & 100\\
      $BWC(-0.8)$ & 94 &100  & 100 & 100 & 100 & 100 & 100 & 100 & 100 & 100  & 100 \\
     
     $BCvM(0.5)$ & 32 & 25 & 24 & 28 & 31 & 33 & 34 & 34 & 34 & 25 & 33 \\
     $BCvM(1)$ & 80 & 73 & 68 & 76 & 81 & 83 & 84 & 86 & 86 & 70 & 84 \\
     $BCvM(1.5)$ & 98 & 95 & 92 & 95 & 97 & 98 & 98 & 99 & 100 & 93 & 98 \\
     $BCvM(2)$ & 100 & 99 & 99 & 99 & 100 & 100 & 100 & 100 & 100 & 99 & 100 \\
     $BCvM(3)$ & 100 & 100 & 100 & 100 & 100 & 100 & 100 & 100 & 100 & 100  & 100\\
\end{tabular}}
 \caption{Empirical sizes and powers of tests in Case I ($n=50$, $\alpha=0.05$)}
    \label{tab: circular50}
\end{table}

\subsection{Case II: spherical-spherical data}

In this case, due to the lack of existing tests for this kind of data, we compare the power performance of our tests among themselves against mixtures of von Mises-Fisher distributions (denoted by $vMF(\pmb{\mu},\kappa)$) whose  density is given by
 \begin{align*}
            f(\pmb{x})=\frac{\kappa^{\frac{d}{2}-1}}{(2\pi)^\frac{d}{2}I_{\frac{p}{2}-1}(\kappa)}e^{\kappa\pmb{\mu}^T\pmb{x}},\pmb{x}\in \mathcal{S}^{d-1},||\pmb{\mu}||=1,\kappa\geq0.
        \end{align*}
The results are presented in Table \ref{tab: spherical20}  and  \ref{tab: spherical50}. We can see that the tests are well-calibrated. Also interesting is that the impact of $a$ against this class of alternatives is smaller than in Tables \ref{tab: circular20}  and \ref{tab: circular50}. As might not be too surprising, the impact of $p$ diminish the impact of components distribution.

    \begin{table}[t]
\scalebox{0.7}{
%\begin{table}[!ht]
    \centering
\begin{tabular}{c|ccccccccccc}
 %  &\multicolumn{11}{c}{$n=20$} %& \multicolumn{8}{c}{$n=50$}    \\
     Model  &  $D_{0.25}$ & $D_{0.5}$ & $D_{0.75}$ & $D_{1}$ & $D_{1.25}$ 
     & $D_{1.5}$ & $D_{1.75}$& $D_k$ & $D_l$      \\ \hline
     $vMF((1,0,0),0)\times vMF((1,0,0),0)$ & 5 & 5 & 5 & 5 & 5 & 5 & 5 & 5 & 5\\
     $vMF((1,0,0,0),0)\times vMF((1,0,0,0),0)$ & 6 & 6 & 6 & 6 & 6 & 6 & 6 & 6 & 6 \\\hline

$vMF((1,0,0),0)\times Mix(vMF((1,0,0),0),vMF((1,0,0),2),0.25)$ & 20 & 21 & 21 & 21 & 22 & 22 & 22 & 20 & 21\\
    $vMF((1,0,0),0)\times Mix(vMF((1,0,0),0),vMF((1,0,0),2),0.5)$ &  75 & 76 & 77 & 77 & 77 & 77 & 78 & 75 & 77\\

    $vMF((1,0,0,0),0)\times Mix(vMF((1,0,0,0),0),vMF((1,0,0,0),2),0.25)$ & 21 & 21 & 22 & 22 & 22 & 22 & 22 & 22 & 22 \\

    $vMF((1,0,0,0),0)\times Mix(vMF((1,0,0,0),0),vMF((1,0,0,0),2),0.5)$ & 75 &  76 & 76 & 76 & 77 & 77 & 78 & 75 & 77 \\

   %  $vMF((0.5,0.5,0.5,0.5),0)\times Mix(vMF((0.5,0.5,0.5,0.5),0),vMF((1,0,0,0),2),0.25)$ & 0.1980 & 0.2050 & 0.2095 &0.2115 & 0.2155 & 0.2205 & 0.2230\\
   %  $vMF((0.5,0.5,0.5,0.5),0)\times Mix(vMF((0.5,0.5,0.5,0.5),0),vMF((1,0,0,0),2),0.5)$ \\
\end{tabular}}
    \caption{ Empirical sizes and powers of the tests in Case II ($n=20$, $\alpha=0.05$)}
    \label{tab: spherical20}
\end{table}

\begin{table}[]
\scalebox{0.7}{
%\begin{table}[!ht]
    \centering
\begin{tabular}{c|ccccccccccc}
 %  &\multicolumn{11}{c}{$n=20$} %& \multicolumn{8}{c}{$n=50$}    \\
     Model  &  $D_{0.25}$ & $D_{0.5}$ & $D_{0.75}$ & $D_{1}$ & $D_{1.25}$ 
     & $D_{1.5}$ & $D_{1.75}$& $D_k$ & $D_l$      \\ \hline
$vMF((1,0,0),0)\times vMF((1,0,0),0)$ & 5 & 5 & 4 & 4 & 4 & 4 & 4 & 5 & 4\\
 $vMF((1,0,0,0),0)\times vMF((1,0,0,0),0)$ & 6 & 5 & 5 & 5 & 5 & 5 & 4 & 6 & 5\\\hline
   $vMF((1,0,0),0)\times Mix(vMF((1,0,0),0),vMF((1,0,0),2),0.25)$ & 56 & 57 & 57 & 57 & 58 & 59 & 60 & 55 & 58 \\
    $vMF((1,0,0),0)\times Mix(vMF((1,0,0),0),vMF((1,0,0),2),0.5)$ & 100 & 100 & 100 &  100 & 100 & 100 & 100 & 100 & 100\\    
     $vMF((1,0,0,0),0)\times Mix(vMF((1,0,0,0),0),vMF((1,0,0,0),2),0.25)$ & 54 & 55 & 56 & 56 & 56 & 57 & 57 & 54 & 56 \\

    $vMF((1,0,0,0),0)\times Mix(vMF((1,0,0,0),0),vMF((1,0,0,0),2),0.5)$ & 100 & 100 & 100 & 99 & 99 & 99 & 99 & 100 & 99\\

\end{tabular}}
    \caption{ Empirical sizes and powers in Case II ($n=50$, $\alpha=0.05$)}
    \label{tab: spherical50}
\end{table}

\begin{comment}
\begin{table}[]
    \centering

\scalebox{0.7}{
\begin{tabular}{c|ccccccccccc}
 %  &\multicolumn{11}{c}{$n=20$} %& \multicolumn{8}{c}{$n=50$}    \\
     Model  &  $D_{0.25}$ & $D_{0.5}$ & $D_{0.75}$ & $D_{1}$ & $D_{1.25}$ 
     & $D_{1.5}$ & $D_{1.75}$& $D_k$ & $D_l$      \\ \hline

$VMF((1,0,0),0)\times VMF((1,0,0),0)$ &0.0550 & 0.0520 & 0.0425 & 0.0395 & 0.0375 & 0.0365 & 0.0390 &  0.0510 & 0.0385\\

   $Mix(VMF((1,0,0),0),VMF((1,0,0),2),0.25)$ & 0.5600 & 0.5665 & 0.5715 & 0.5740 & 0.5830 & 0.5930 & 0.5980 & 0.5515 & 0.5775 \\
    $Mix(VMF((1,0,0),0),VMF((1,0,0),2),0.5)$ & 0.9965 & 0.9965 & 0.9960 &  0.9955 & 0.9950 & 0.9950 & 0.9955 & 0.9965 & 0.9950\\

\end{tabular}}
    \caption{ Empirical sizes and powers $n=50$, $\alpha=0.05$}
    \label{tab: spherical50}
\end{table}
\end{comment}

\subsection{Case III: circular-linear case}
We compare the power of our tests with the kernel density estimator based from \cite{garcia2014test}  given by

%test based on directional kernel density estimation proposed in {\color{mediumslateblue} Garc\'ia-Portugu\'es et al. (2014)} with test statistic
        \begin{align*}
            T_n&=\pmb{I}_n\Big(\frac{1}{n^2}\pmb{\Psi}(h)\circ\pmb{\Omega}(g)-\frac{2}{n^2}\pmb{\Psi}(h)\pmb{\Omega}(g)+\frac{1}{n^4}\pmb{\Psi}(h)\pmb{I}_n\pmb{I}_n^T\pmb{\Omega}(g)\Big)\pmb{I}_n^T,
        \end{align*}
        where $\circ$ denotes the Hadamard product, $h$ and $g$ are bandwidths, and $\pmb{\Psi}(h)$ and $\pmb{\Omega}(g)$ are $n\times n$ matrices given by
        \begin{align*}
            \pmb{\Psi}(h)=\left(\frac{C_q({1}/{h^2})^2}{C_q(||\pmb{X_i}+\pmb{X_j}||/h^2)}\right)_{ij},\quad \pmb{\Omega}(g)=\Big(\phi_{\sqrt{2g}}(Z_i-Z_j)\Big)_{ij},
        \end{align*}
        where $\pmb{I}_n$ is a vector of $n$ ones and $C_q=\kappa^\frac{q-1}{2}\Big((2\pi)^{\frac{q+1}{2}}I_{\frac{q-1}{2}}(\kappa)\Big)^{-1}, \kappa\geq 0, $ and $I_{\nu}$ is the modified Bessel function of order $\nu$.
Here, as recommended in \cite{garcia2014test}, the bandwidth parameters are estimated by bootstrap likelihood cross-validation, which considerably slows down the computation of test statistics. For this reason, in our power study, instead of bootstrap algorithm \ref{bootstrap} we use its warp-speed modification. 

The alternatives we consider are
\begin{itemize}
    \item generated with the von Mises copula $VMC(k)$ and uniform marginals;

    \item projected normal-based $PN(d,\Sigma)$  obtained from $d+1$ dimensional vector from $d+1$-dimensional normal distribution  by projecting the first $d$ components on the sphere, i.e. using the following set of transformations $(X_{1},...,X_{d},Y)\sim \mathcal{N}_{d+1}(0,\Sigma)	\longrightarrow(\underbrace{{X_{1}/||\pmb{X}||},...,X_{d}/||\pmb{X}||,}_{\in \mathcal{S}_{d-1}}Y)$.
\end{itemize}

The results are presented in Tables \ref{tab: circLinBrzo20} and \ref{tab: circLinBrzo50}. As before, the tests are all well calibrated. However, now the impact of the kernel is very important and an outlook for future study might be to think about a data driven selection of the kernel. It is also interesting to note that $D_k$ shows a similar power performance as $D_a$ for smaller $a$, while $D_l$'s behaviour is more closer to the one of $D_{a}$ with larger $a$.  In addition, the powers of $D_{0.25}$ are similar to those of $T$. In view of computation time $D_{0.25}$ might then be the favorable choice for a practitioner.

\begin{table}[t]
    \centering

\scalebox{0.7}{
\begin{tabular}{c|ccccccccccc}
 %  &\multicolumn{11}{c}{$n=20$} %& \multicolumn{8}{c}{$n=50$}    \\
     Model  & $T_n$ & $D_{0.25}$ & $D_{0.5}$ & $D_{0.75}$ & $D_{1}$ & $D_{1.25}$ 
     & $D_{1.5}$ & $D_{1.75}$& $D_k$ & $D_l$      \\ \hline
   $VMC(0)$ & 6 & 6 & 5 & 5 & 5 & 5 & 5 & 5 & 5 & 5\\
   
$PN(2,(0.1,0,0))%=\begin{pmatrix}1&0.1&0\\0.1&1&0\\0&0&1\end{pmatrix}
$ & 5 & 5 & 6 & 6 & 6 & 5 & 5 & 5 & 5 & 6\\\hline 
$VMC(0.5)$ & 11 & 10 & 11 & 11 & 10 & 10 & 10 & 9 & 11 & 9\\
$VMC(1)$ & 38 & 36 & 38 & 32 & 28 & 23 & 19 & 17 & 38 & 17 \\
$VMC(2)$ & 91 & 90 & 89 & 84 & 75 & 67 & 56 & 47 & 89 & 49 \\

$PN(2,(0.1,0.5,0.3))%\Sigma=\begin{pmatrix}    1&0.1&0.5\\0.1&1&0.3\\0.5&0.3&1 \end{pmatrix}
$ & 26 & 28 & 40 & 45 & 47 & 49 & 49 & 50 & 33 & 46\\

$PN(2,(0.1,0.8,0.3))%\Sigma=\begin{pmatrix}  1&0.1&0.8\\0.1&1&0.3\\0.8&0.3&1\end{pmatrix}
$ & 78 & 76 & 87 & 91 & 95 & 95 & 96 & 96 & 80 & 92\\

$PN(2,(0.1,0.9,0.3))%\Sigma=\begin{pmatrix}    1&0.1&0.9\\0.1&1&0.3\\0.9&0.3&1\end{pmatrix}
$ & 96 & 95 & 98 & 99 & 99 & 99 & 99 & 99 & 97 & 99 
\end{tabular}}
    \caption{ Empirical sizes and powers in Case III ($n=20$, $\alpha=0.05$)}
    \label{tab: circLinBrzo20}
\end{table}

\begin{table}[t]
    \centering

\scalebox{0.7}{
\begin{tabular}{c|ccccccccccc}
 %  &\multicolumn{11}{c}{$n=20$} %& \multicolumn{8}{c}{$n=50$}    \\
     Model  & $T_n$ & $D_{0.25}$ & $D_{0.5}$ & $D_{0.75}$ & $D_{1}$ & $D_{1.25}$ 
     & $D_{1.5}$ & $D_{1.75}$& $D_k$ & $D_l$      \\ \hline
   $VMC(0)$ & 5 & 5 & 4 & 5 & 5 & 5 & 5 & 5 & 4 & 5\\
   $\Sigma=\begin{pmatrix}
    1&0.1&0\\0.1&1&0\\0&0&1
\end{pmatrix}$ & 5 & 5 & 5 & 5 & 5 & 5 & 5 & 5 &  6 & 5\\\hline
$VMC(0.5)$ & 32 & 32 & 32 & 30 & 26 & 24 & 22 & 19 & 33 & 20 \\
$VMC(1)$ & 85 & 84 & 84 & 78 & 73 & 66 & 58 & 52 & 84 & 54 \\
$VMC(2)$ & 100 & 100 & 100 & 100 & 100 & 99 & 98 & 96 & 100 & 96 \\
$\Sigma=\begin{pmatrix}
    1&0.1&0.5\\0.1&1&0.3\\0.5&0.3&1
\end{pmatrix}$ & 73 & 80 & 89 & 91 & 93 & 93 & 94 & 94 & 81 & 92 \\

$\Sigma=\begin{pmatrix}
    1&0.1&0.8\\0.1&1&0.3\\0.8&0.3&1
\end{pmatrix}$ & 100 & 100 & 100 & 100 & 100 & 100 & 100 & 100 & 100 & 100\\

$\Sigma=\begin{pmatrix}
    1&0.1&0.9\\0.1&1&0.3\\0.9&0.3&1
\end{pmatrix}$ & 100 & 100 & 100 & 100 & 100 & 100 & 100 & 100 & 100 & 100\\
\end{tabular}}
    \caption{ Empirical sizes and powers in Case III ($n=50$, $\alpha=0.05$)}
    \label{tab: circLinBrzo50}
\end{table}

\section{Real data example}\label{sec:real}

This section provides three real data examples which clearly illustrate the applicability of the proposed methodology. 

\textbf{Example 1:} In a medical experiment, various measurements were taken on 10 medical 
students several times daily for a period of several weeks \cite{downs1974rotational}. 
The estimated peak times (converted into angles $\theta$ and $\phi$) for two successive 
measurements of diastolic blood pressure are given in Table \ref{tab: datasetBloodPreasure}.
\begin{table}[!h]
\centering
\begin{tabular}{c|cccccccccc}
$\boldsymbol{\theta}$ & 30 & 15 & 11 & 4 & 348 & 347 & 341 & 333 & 332 & 285\\
$\boldsymbol{\phi}$ & 25 & 5 & 349 & 358 & 340 & 347 & 345 & 331 & 329 & 287\\
\end{tabular}
\caption{Dataset 1 - Blood preasure}
\label{tab: datasetBloodPreasure}
\end{table}
Although the sample size is small, all considered test statistics reject the hypothesis of independence (all the p-values are smaller than $10^{-3}$).  
%The results, presented in Table \ref{tab: p-values} indicates that the assumption of independence between two successive daily peaks of blood pressure, is very realistic.

%\textbf{Example 2:} This data set contains 38 phase or peak expression times of
%synchronized circadian-related genes common to heart and liver tissue in vivo \cite{liu2006phase}. The  phase angles (in radians) of circadian-related transcripts in heart and liver are given in Table \ref{tab: datasetPeakTimes}. It can be seen that all tests support the hypothesis of dependence (see Table \ref{tab: p-values}). 

%\begin{table}[!htb]
%\centering
%\begin{tabular}{c|cccccccccccc}
%$\boldsymbol{\theta}$ & 0.12 & 0.27 & 0.29 %& 0.3 & 0.31 & 0.34 & 0.35 & 0.58 & 0.62 & %1.6 \\ 
%$\boldsymbol{\phi}$ & 0.61 & 0.95 & -2.85 & %0.67 & -0.13 & 0.08 & 2.67 & 1.72 & 1.45 & %1.59 \\\hline
%$\boldsymbol{\theta}$ & 2.35 & 2.62  & 2.83 %& -3.06 & -2.86 & -2.77 & -2.69 & -2.57 & %-2.56 & -2.45 \\
%$\boldsymbol{\phi}$ & -2.51  & -2.92 & 1.42 %& 2.74 & 2.88 & -3.01 & -2.69 & 3.05 & %-2.35 & 2.68 \\\hline
%$\boldsymbol{\theta}$ & -2.43 & -2.37  & %-2.18 & -2.16 & -2.04 & -1.61 & -1.32 & %-1.22 & -0.84 & -0.77 \\ 
%$\boldsymbol{\phi}$ & -2.86 & -2.51 & 2.69 %& -2.11 & -1.48 & -2.06 & -2.63 & -1.49 & %-0.83 & 0.86 \\\hline
%$\boldsymbol{\theta}$ &-0.38 & -0.36 & %-0.26 & -0.19 & -0.18 & -0.13 & -0.12 & %-0.025\\
%$\boldsymbol{\phi}$ & 0.26 & 1.5& 1.03  & %0.33 & -1.15 & -0.21 & -0.55 & 0.91\\
%\end{tabular}
%\caption{Dataset 2 - Peak expression times}
%\label{tab: datasetPeakTimes}
%\end{table}

\textbf{Example 2:} We consider the wind direction at 6 a.m. and 12 p.m.  measured each day at a weather station in Milwaukee for 21 consecutive days (see \cite{johnson1977measures}, and  Table \ref{tab: datasetWind} for the data set). 
Although none of the test statistics reject the hypothesis for the level of significance 0.05, for the level of 0.1 all novel test statistics show the ability to detect a discrepancy from the independence assumption.

\begin{table}[!h]
\centering
\begin{tabular}{ c|ccccccccccc  }
\textbf{6 a.m.} & 356 & 97.2 & 211 & 232 & 343 & 292 & 157 &  302 & 335 & 302 & 324\\
\textbf{12 p.m.} & 119 & 162 & 221 & 259 & 270 & 28.8 & 97.2 & 292 & 39.6 & 313 &94.2\\\hline
 \textbf{6 a.m.}  & 84.6 & 324 & 340 & 157 & 238 & 254 & 146 & 232 & 122 & 329 \\

 \textbf{12 p.m.} & 45 & 47 & 108 & 221 & 270 & 119 & 248 & 270 & 45 & 23.4
\end{tabular}
\caption{Dataset 2 - Wind direction}
\label{tab: datasetWind}
\end{table}

\begin{table}[!h]
    \centering
    \scalebox{0.9}{
   \begin{tabular}{c|ccccccccccc}
 %  &\multicolumn{11}{c}{$n=20$} %& \multicolumn{8}{c}{$n=50$}    \\
       & $C$ & $T_1$ & $D_{0.25}$ & $D_{0.5}$ & $D_{0.75}$ & $D_{1}$ & $D_{1.25}$ 
     & $D_{1.5}$ & $D_{1.75}$& $D_k$ & $D_l$ %& C & $D_{0.25}$ & $D_{0.5}$ & $D_{0.75}$ & $D_{1}$ & $D_{1.25}$      & $D_{1.5}$ & $D_{1.75}$
     \\ \hline
      Dataset 2 & 0.225 & 0.087 & 0.069 & 0.057 & 0.064 & 0.072 & 0.07 & 0.073 & 0.071 & 0.065 & 0.073  \\
     % Dataset 4 &  &  & 2 & 2 & 2 & 2 & 2 & 2 & 2 & 2 & 2
      \end{tabular}}
    \caption{Data set 2: $p-$values}
    \label{tab: p-values2}
\end{table}

\textbf{Example 3:} We consider magnetic remanence at 680°C and 685°C  in each of 52 rock specimens was measured (see \cite{fisher1986correlation}). For all kernels, test statistic rejected the null hypothesis (see Table \ref{tab: p-values3}) which is in concordance with results presented in original work.

\begin{table}[!h]
    \centering
    \scalebox{0.9}{
   \begin{tabular}{c|ccccccccc}
 %  &\multicolumn{11}{c}{$n=20$} %& \multicolumn{8}{c}{$n=50$}    \\
        & $D_{0.25}$ & $D_{0.5}$ & $D_{0.75}$ & $D_{1}$ & $D_{1.25}$ 
     & $D_{1.5}$ & $D_{1.75}$& $D_k$ & $D_l$ %& C & $D_{0.25}$ & $D_{0.5}$ & $D_{0.75}$ & $D_{1}$ & $D_{1.25}$      & $D_{1.5}$ & $D_{1.75}$
     \\ \hline
      Dataset 3   & 0.02 & 0.02 & 0.02 & 0.019 & 0.017 & 0.018 & 0.018 & 0.021 & 0.015
      \end{tabular}}
    \caption{Data set 3: $p-$values}
    \label{tab: p-values3}
\end{table}

%\begin{table}[]
 %   \centering
  %  \scalebox{0.9}{
  % \begin{tabular}{c|ccccccccccc}
 %  &\multicolumn{11}{c}{$n=20$} %& \multicolumn{8}{c}{$n=50$}    \\
   %    & $C$ & $T_1$ & $D_{0.25}$ & %$D_{0.5}$ & $D_{0.75}$ & $D_{1}$ & %$D_{1.25}$ 
    % & $D_{1.5}$ & $D_{1.75}$& $D_k$ & %$D_l$ %& C & $D_{0.25}$ & $D_{0.5}$ & %$D_{0.75}$ & $D_{1}$ & $D_{1.25}$      %& $D_{1.5}$ & $D_{1.75}$
    % \\ \hline
    % Dataset 1  & 0 & 0 & 0 & 0 & 0 & 0 & 0 & 0& 0 & 0  & 0 \\
    %  Dataset 2 & 0 & 0 & 0 & 0 & 0 & 0 & 0 %& 0 & 0 & 0 & 0 \\
    %  Dataset 3 & 0.225 & 0.087 & 0.069 & 0.057 & 0.064 & 0.072 & 0.07 & 0.073 & 0.071 & 0.065 & 0.073  \\
    %  Dataset 4 &  &  & 0.02 & 0.02 & 0.02 & 0.019 & 0.017 & 0.018 & 0.018 & 0.021 & 0.015
    %  \end{tabular}}
    %\caption{$p-$values}
    %\label{tab: p-values}
%\end{table}

\section{Concluding remarks and outlook }

In this study, we revisit the energy-based test for independence and consider its kernel-based modification tailored for directional data.  We delve into the detailed exploration of the properties within this test class, leading to the conclusion that the incorporation of arbitrary strongly negative-definite kernels significantly enhances the test's flexibility. A noteworthy advantage of this approach is its applicability to various types of directional data, allowing tests of independence among vectors of different types.

Furthermore, the computational aspect of these tests is highlighted as advantageous. Unlike some competitors, the computation of statistics does not necessitate the estimation of hyperparameters nor involve numerical integration. This cost-effectiveness becomes particularly relevant when dealing with large sample sizes and higher dimensions.

One of the natural directions of further research is to exchange the class of kernels that is used in a way to consider also kernels that are not functions of the distance of the arguments. Another interesting research option is to develop a data-driven choice of kernels. For the latter, different asymptotic properties are expected.

Given the computational efficiency of the tests under consideration, an additional direction for exploration involves applying these tests to variable selection problems (see e.g. \cite{fan2008sure} and \cite{edelmann2020marginal}). Such an extension has the potential to greatly expand the range of situations in which these tests can be effectively applied.

\section*{Acknowledgements} {We are thankful to Eduardo
García-Portugués for kindly providing a code for the computation of statistic $T_n$ introduced in \cite{garcia2014test}.

The research was supported by the bilateral cooperation project ”Modeling complex data - Selection and Specification” between the Federal Republic of Germany and the Republic of Serbia. The project underlying this article was funded by the Federal Ministry of Education and Research of Germany. The work of B. Milošević and M. Cuparić is supported by the Ministry of Science, Technological Development and Innovations of the Republic of Serbia (the contract 451-03-47/2023-01/200104) and also supported by the COST action CA21163 - Text, functional and other high-dimensional data in econometrics: New models, methods, applications (HiTEc).}
\bibliographystyle{abbrv}
%\bibliography{LiteratureCircular}

\begin{appendix}
\section{Kernel-induced distance of probability measures}\label{app:Kernel}
    In this paper we consider distance covariance functions induced by kernels for spherical data. For this task, we need some preliminary statements taken with slight modifications from \cite{R:2013}, Section 22.

Let  $(\mathcal{X},\mathcal{A})$ be a measurable space and denote by $\mathcal{M}$ the set of probability measures on $(\mathcal{X},\mathcal{A})$. A map $K:\mathcal{X}^2\rightarrow\mathbb{R}$ is called a \textit{negative definite} kernel function if for any $n\in\N$, any arbitrary set of real values $a_1,\ldots,a_n$ and any aribtrary set $x_1,\ldots,x_n$ of points in $\mathcal{X}$ the following holds:
\begin{equation}\label{eq:negdef}
\sum_{j,k=1}^na_ja_kK(x_j,x_k)\le0.
\end{equation}
We call $K$ \textit{strictly} negative definite if the equality in \eqref{eq:negdef} holds only for the case $a_1=a_2=\cdots=a_n=0$. Let $Q\in\mathcal{M}$ and $h$ be a function integrable w.r.t. $Q$ such that 
\begin{equation*}
	\int_{\mathcal{X}}h(x)\,\mbox{d}Q(x)=0.
\end{equation*} 
We define $K$ to be a \textit{strongly} negative definite kernel if $K$ is negative definite and 
\begin{equation*}
	\int_{\mathcal{X}}\int_{\mathcal{X}}K(x,y)h(x)h(y)\,\mbox{d}Q(x)\mbox{d}Q(y)=0.
\end{equation*} 
implies that $h\equiv0$ $Q$-almost everywhere for all $Q\in\mathcal{M}$. Denote by $\mathcal{M}_K\subset\mathcal{M}$ the subset of probability measures for which
\begin{equation*}
	\int_{\mathcal{X}}K(x,x)\mbox{d}\mu(x)<\infty.
\end{equation*}
Next, define 
\begin{eqnarray}
 N_K(\mu,\nu)&=&2\int_{\mathcal{X}}\int_{\mathcal{X}}K(x,y)\mbox{d}\mu(x)\mbox{d}\nu(y)-\int_{\mathcal{X}}\int_{\mathcal{X}}K(x,y)\mbox{d}\mu(x)\mbox{d}\mu(y)-\int_{\mathcal{X}}\int_{\mathcal{X}}K(x,y)\mbox{d}\nu(x)\mbox{d}\nu(y).\label{eq:Ndist}
\end{eqnarray}
for $\mu,\nu\in\mathcal{M}_K$. With these preliminary statements, we can restate Theorem 22.2.1 from \cite{R:2013}.

\begin{theorem}
	Let $K$ be a strongly negative definite kernel on $\mathcal{X}^2$ satisfying 
	\begin{equation*}
		K(x,y)=K(y,x),\quad\mbox{and}\quad K(x,x)=0\,\mbox{ for all}\; x,y\in\mathcal{X}.
	\end{equation*}
Then $N_K^{1/2}$ is a distance on $\mathcal{M}_K$.
\end{theorem}
Let $X,Y$ be independent random variables having the CDFs related to $\mu,\nu\in\mathcal{M}_K$. Denote by $X',Y'$ independent copies of $X,Y$ and assume that $X,X',Y,Y'$ are mutually independent. Then \eqref{eq:Ndist} takes the form
\begin{eqnarray}
	N_K(\mu,\nu)&=&2EK(X,Y)-EK(X,X')-EK(Y,Y').
\end{eqnarray}
Note that $N_K=0$ implies equality in distribution of $X$ and $Y$. A famous example is the special case $\mathcal{X}=\R^d$ and $K(x,y)=\|x-y\|^r$ for some $r\in(0,2)$, since then we have the (generalized) energy distance. %Of course one can consider other measurable spaces and find numerous of strongly negative definite kernels. We will detail interesting special cases later.

%In what follows, consider two random elements $X,Y$ taking values in measurable spaces $(\mathcal{X}_1,\mathcal{A}_1)$ and $(\mathcal{X}_2,\mathcal{A}_2)$ respectively. We consider the random element $(X,Y)$ taking values in $\mathcal{X}_1\times \mathcal{X}_2$ and \textcolor{red}{We have to define distance covariance here. I actually don't see how to define it with general kernels. I am working on it.}\\[3mm]
\end{appendix}

\end{document}